# Anatomy of spin Hall effect in ferromagnetic metals

Fanxing Zheng[1], Jianting Dong[1], Xinlu Li[1], Meng Zhu[1], Ye Zhou[1], and Jia Zhang[1*]

[1]School of Physics and Wuhan National High Magnetic Field Center,

Huazhong University of Science and Technology, 430074 Wuhan, China.

* jiazhang@hust.edu.cn

## Abstract

The spin Hall effect in nonmagnetic materials has been intensively studied and became one of the most crucial spin-charge conversion mechanism in spintronics. However, the spin Hall effect in ferromagnetic metals has been less investigated and remains unclear. In this work, we investigate the spin Hall effect in representative ferromagnetic alloy by using first-principles calculations. We first clarify the spin Hall effect into three different types including conventional (CSHE), spin anomalous (SAHE) and magnetic spin Hall effect (MSHE) and then calculate the corresponding spin Hall conductivity and spin Hall angle for (Fe, Co, Ni)Pt, NiFe and CoFe alloy. We find the above three spin Hall mechanisms do coexist in ferromagnetic metals. Particularly, for Pt-based ferromagnetic alloy, a sizable conventional and magnetic spin Hall angles comparable to that of Pt have been predicted. The remarkable unconventional spin Hall effect in ferromagnetic metal may enrich the spin-charge conversion phenomena. For instance, the spin current generated by remarkable MSHE with out-of-plane spin-polarization should be helpful for field-free switching of perpendicular magnetization through spin-orbit torque effect. This work may stimulate future studies on the spin Hall effect in ferromagnetic metals and pave their promising applications for spin-charge conversion devices in spintronics.

## Introduction

In nonmagnetic materials, the spin Hall effect (SHE) originates from the asymmetric deflection of electrons with opposite spin in the presence of spin-orbit coupling when charge flows[1]. A transverse spin current with the spin-polarization perpendicular to the plane of charge and spin current will be generated by SHE, which can be described as $\boldsymbol{J}_s = \theta_{SH}(\hbar/2e)\boldsymbol{J}_c \times \hat{\boldsymbol{\sigma}}$, where $\theta_{SH}$ is the spin Hall angle, $\boldsymbol{J}_c$ is charge current density, $\hat{\boldsymbol{\sigma}}$ is the spin-polarization direction, $\hbar$ is the reduced Planck constant and $e$ is the elementary charge. SHE and inverse SHE (ISHE) in nonmagnetic heavy metals like Pt, Ta, W and alloy have been intensively studied for spin-charge conversion in spintronics[2]. Especially, in "heavy metal/ferromagnet (FM)" bilayer, a charge current in heavy metal produces spin current by SHE mechanism and generates spin-orbit torque (SOT) on adjacent ferromagnets (SHE-SOT), leading to current-induced switching of magnetization[3][3][5]. The SHE-SOT provides an efficient writing scheme for the next-generation magnetic random access memory (SOT-MRAM)[6].

Unlike the discovery of ordinary Hall effect in nonmagnetic materials and anomalous Hall effect (AHE) in ferromagnets[7], The SHE in ferromagnetic metals has been overlooked for quite a long time. Only in recent years, a few research refer to noteworthy spin Hall effect in ferromagnets. Certainly, there is remarkable SHE in ferromagnetic metals sharing the same physical origin as in nonmagnetic metals (In the following section, we call it conventional spin Hall effect, denoted by CSHE). Conventional inverse spin Hall was first experimentally discovered in permalloy[8], which was found to be magnetization independent. The spin current associated with AHE in ferromagnets was proposed later, known as spin anomalous Hall effect (SAHE)[9]. The existence of SAHE in ferromagnets requires the noncollinear alignment of magnetization direction $\hat{m}$ and charge current $\boldsymbol{J}_c$ with the spin current propagating along $\hat{m} \times \boldsymbol{J}_c$ and the spin-polarization $\boldsymbol{\sigma}$ parallel to magnetization direction. There are subsequent experiments demonstrating that in "FM (bottom)/non-magnetic metal (NM)/FM (top)" sandwich structures, electric current in the bottom ferromagnet induces spin current by SAHE and exerts a spin-orbit torque on top ferromagnet[10][11]. If strong enough, the SAHE-SOT will be able to switch the magnetization [12][13][14][15]. Recently, SAHE in ferromagnets Fe, Co, Ni and

L10-(Fe, Co, Ni)Pt have been investigated by first-principles calculation[16][17].

The CSHE and SAHE is time reversal even (*T-even*) and the corresponding spin Hall conductivity (SHC) does not change sign with respecting to the reversal of magnetization. Imposed by symmetry of magnetic point group[18], there is also the third type of spin Hall effect called magnetic spin Hall effect (MSHE) in ferromagnetic metal and the corresponding spin Hall conductivity is time reversal odd (*T-odd*)[19][20][21], which should be vanishing for nonmagnetic materials. The MSHE has been first experimentally observed in non-collinear anitiferromagnet $Mn_3Sn$[22], and recently been theoretically investigated in Fe, Co, Ni[20].

However, the magnitude of spin Hall angle, and the features of spin current generated by different spin Hall mechanisms in practical ferromagnetic metals remain obscure. In this work, by using first-principles calculations we investigate the spin Hall effect in representative ferromagnetic $(Fe, Co, Ni)_{50}Pt_{50}$ and other alloys. We first clarify and analyze the three different SHE mechanisms in ferromagnetic metal. And then the spin Hall conductivity and spin Hall angle are evaluated based on the Kubo-Bastin linear response formalism. Our results indicate that the Pt-based ferromagnetic alloys have a large conventional spin Hall angle $\theta_{CSH}$, comparable to that in Pt. What's more, Pt-based ferromagnetic alloy has remarkable MSHE with magnetic spin Hall angle $\theta_{MSH}$ around several percent, which is at the same order of CSHE. We also show that the magnitude of the CSHE, SAHE and MSHE can be optimized by varying alloy composition. We discuss the fascinating applications of spin current generated by spin Hall effect in ferromagnetic metals for instance for the field-free switching of perpendicular magnetization by SOT.

## Calculation method

The first-principles calculations are performed by using fully relativistic Korringa–Kohn–Rostoker (KKR) Green's function method[23][24]. The cutoff of *lmax*=3 is used for the angular momentum expansion of Green's function, and the Vosko-Wilk-Nussair (VWN) type of LDA exchange and correction potential was employed[25]. The energy integration has been performed on a semicircle in the complex plane by using 50 energy points and $45\times45\times45$ $\boldsymbol{k}$-points in the Brillouin zone (BZ) to obtain self-consistent field (SCF) potentials. The chemical disorder of alloy has been treated within the framework of coherent potential approximation (CPA) [26][27].

In Cartesian coordinate, the linearly responded spin current propagates along $\mu$ axis with spin-polarization in $\xi$ direction stimulated by electric field in $\nu$ axis can be expressed as:

$$J_\mu^\xi = \sum_\nu J_{\mu\nu}^\xi = \sum_\nu \sigma_{\mu\nu}^\xi E_\nu \qquad (\mu,\nu,\xi \in \{x, y, z\}) \qquad (1)$$

The spin Hall conductivity $\sigma_{\mu\nu}^\xi$ is a tensor of rank three, and it can be evaluated based on the Kubo-Bastin formalism by dividing it into Fermi surface and Fermi sea contributions [28]. For alloy system, the SHCs calculated by this method with vertex correction include both intrinsic and extrinsic contributions (*e.g.* skew scattering and side jump). The detailed descriptions on calculation of spin Hall conductivity can be found in Supplementary Note1. In this work, approximately $10^7$ $\boldsymbol{k}$ points in the BZ are found to be sufficient for convergence of SHC.

**Symmetry analysis of spin Hall conductivity tensor**

The A1 phase of (Fe, Co, Ni)$_{50}$Pt$_{50}$ alloy in fcc crystal structure belongs to the magnetic Laue group 4/mm'm'. The direction of magnetic moment $\boldsymbol{M}$ has been set to be along $\hat{x}$ direction in our calculation. Thus, the symmetry imposed spin Hall conductivity tensor with spin-polarization $\boldsymbol{\sigma}$ of spin current along $\hat{x}$, $\hat{y}$, $\hat{z}$ axis can be written as [18]:

$$\sigma_{\mu\nu}^x = \begin{pmatrix} \sigma_{xx}^x & 0 & 0 \\ 0 & \sigma_{yy}^x & \sigma_{yz}^x \\ 0 & \sigma_{zy}^x & \sigma_{zz}^x \end{pmatrix} \quad \sigma_{\mu\nu}^y = \begin{pmatrix} 0 & \sigma_{xy}^y & \sigma_{xz}^y \\ \sigma_{yx}^y & 0 & 0 \\ \sigma_{zx}^y & 0 & 0 \end{pmatrix} \quad \sigma_{\mu\nu}^z = \begin{pmatrix} 0 & \sigma_{xy}^z & \sigma_{xz}^z \\ \sigma_{yx}^z & 0 & 0 \\ \sigma_{zx}^z & 0 & 0 \end{pmatrix} \quad (2)$$

The diagonal spin Hall conductivity matrix elements $\sigma_{xx}^x, \sigma_{yy}^x, \sigma_{zz}^x$ are trivial and they describe the longitudinal spin-polarized current in ferromagnetic metals. The off-diagonal elements $\sigma_{\mu\nu}^\xi$ are spin Hall related conductivity and can be classified into three different types based on underlying spin Hall mechanisms.

By taking $Fe_{50}Pt_{50}$ as an example, the classification of SHCs and the schematic diagram of spin current are shown in Table I. The elements of $\sigma_{\mu\nu}^{\xi}$ with Levi-Civita symbol $\varepsilon_{\mu\nu\xi} \neq 0$ indicate the charge current $\mathbf{J}_c$, spin current $\mathbf{J}_s$, and the spin polarization $\boldsymbol{\sigma}$ are mutually orthogonal. Those spin Hall conductivity are even with respecting to time-reversal operation (*T-even*), i.e. $\sigma_{\mu\nu}^{\xi}(\mathbf{M}) = \sigma_{\mu\nu}^{\xi}(-\mathbf{M})$. Among those SHCs, $\sigma_{yx}^{z}, \sigma_{zx}^{y}, \sigma_{xy}^{z}$ and $\sigma_{xz}^{y}$ can be exclusively attributed to CSHE contribution since the magnetic moment $\mathbf{M}$ is collinear with charge current $\mathbf{J}_c$ and there is no contribution from SAHE. Meanwhile, on account of the rotation symmetry operation around *x*-axis as shown by the schematic diagram in Table I, the SHC has the following relations:

$$\sigma_{yx}^{z} = -\sigma_{zx}^{y}, \quad \sigma_{xy}^{z} = -\sigma_{xz}^{y}$$

Please note that as listed in Table I, $\sigma_{yx}^{z}$ and $\sigma_{xy}^{z}$ are not equal in magnitude in ferromagnetic metal because of the lower symmetry in the presence of magnetic moment and spin-orbit coupling.

The other two SHCs $\sigma_{zy}^{x}, \sigma_{yz}^{x}$ contains contributions from CSHE as well as SAHE since now the spin current propagates along $\hat{m} \times \mathbf{J}_c$ direction with its spin-polarization parallel to the magnetic moment direction. Imposed by the symmetry, the two SHCs have the following relation:

$$\sigma_{zy}^{x} = -\sigma_{yz}^{x}$$

It is possible to extract the spin Hall conductivity solely contributed by SAHE mechanism. The SHC corresponding to SAHE can be obtained by subtracting the conventional spin Hall conductivity from the above two SHCs which contain both CSHE and SAHE contributions as follows:

$$\sigma_{zy}^{SAH} = \sigma_{zy}^{SAH+SHC}(\hat{m} \parallel \hat{x}) - \sigma_{zy}^{SHC}(\hat{m} \parallel \hat{y}) = \sigma_{zy}^{x} - \sigma_{xy}^{z}.$$

The remaining four SHC elements $\sigma_{xy}^{y}, \sigma_{yx}^{y}, \sigma_{xz}^{z}, \sigma_{zx}^{z}$ can be attributed to the MSHE. The MSHE related SHCs are odd respecting to time-reversal operation (*T-odd*), i.e.

$\sigma_{\mu\nu}^{\xi}(\boldsymbol{M}) = -\sigma_{\mu\nu}^{\xi}(-\boldsymbol{M})$. Likewise, the SHC has the following relations due to the symmetry restriction:

$$\sigma_{zx}^{z} = \sigma_{yx}^{y}, \sigma_{xy}^{y} = \sigma_{xz}^{z}.$$

It's worthy noting that, in this case, the charge current $\boldsymbol{J}_c$, spin current $\boldsymbol{J}_s$, and spin polarization $\boldsymbol{\sigma}$ are not mutually orthogonal. Instead, two of them can be parallel to each other. This feature will enable us to generate spin current with special spin polarization that can not be acquired by CSHE and SAHE mechanism. For instance, as shown in Table I, the spin current corresponding to SHCs $\sigma_{zx}^{z}(\sigma_{yx}^{y})$ have the spin-polarization parallel to the spin current propagation direction, while the spin current corresponding to SHCs $\sigma_{xz}^{z}(\sigma_{xy}^{y})$ have the spin-polarization parallel to the charge current direction. From this perspective, the spin current generated by MSHE could break the restriction on spin-polarization by CSHE and SAHE mechanism and may extend its potential applications for emergent spin-charge conversion phenomena as we will discuss later.

For ferromagnetic $Fe_{50}Pt_{50}$ alloy, one can observe that the conventional spin Hall (CSHC) and magnetic spin Hall conductivity (MSHC) have similar magnitude, which indicates dominant CSHE and MSHE mechanism coexisting in FePt alloy. As we will discuss later, although the absolute value of CSHC and MSHC for $Fe_{50}Pt_{50}$ is around a quarter of the SHC in Pt (4336 $\hbar/2e(\Omega\cdot cm)^{-1}$ in our calculation), $Fe_{50}Pt_{50}$ has comparable spin Hall angle as Pt since FePt alloy has relatively lower longitudinal conductivity that Pt.

**TABLE I. Classification of SHE in $Fe_{50}Pt_{50}$.** The calculated SHC (unit: $\hbar/2e(\Omega\cdot cm)^{-1}$) for A1 phase $Fe_{50}Pt_{50}$, and the schematic diagram of spin current generated by CSHE, SAHE and MSHE, where the green and blue arrows indicate the directions of charge current $\boldsymbol{J}_c$ and magnetization $\boldsymbol{M}$. The black and red arrows represents the induced the directions of spin current $\boldsymbol{J}_s$ and the spin polarization $\boldsymbol{\sigma}$.

| Classification of SHE | Spin Hall conductivity | | Direction relationship | Schematic diagram |
|---|---|---|---|---|
| **CSHE** (*T-even*) | $\sigma_{yx}^{z}(-\sigma_{zx}^{y})$ | 1195 | $J_c // M$ $\sigma \perp J_s \perp J_c$ | |
| | $\sigma_{xy}^{z}(-\sigma_{xz}^{y})$ | -1073 | $J_c \perp M$ $\sigma \perp J_s \perp J_c$ | |
| **SAHE+CSHE** (*T-even*) | $\sigma_{zy}^{x}(-\sigma_{yz}^{x})$ | 1354 | $J_c \perp M$ $\sigma \perp J_s \perp J_c$ | |
| **MSHE** (*T-odd*) | $\sigma_{zx}^{z}(\sigma_{yx}^{y})$ | 788 | $J_c // M$ $\sigma // J_s \perp J_c$ | |
| | $\sigma_{xy}^{y}(\sigma_{xz}^{z})$ | -657 | $J_c \perp M$ $\sigma // J_c \perp J_s$ | |

In order to obtain an in-depth understanding on spin Hall effect in ferromagnetic metal, we calculate the spin Hall conductivity and evaluate the spin Hall angle for (Fe, Co, Ni)$_{50}$Pt$_{50}$, Ni$_{80}$Fe$_{20}$ and Co$_{50}$Fe$_{50}$ alloy. The SHCs and the longitudinal conductivity $\sigma_{xx}$ for each alloy can be found in supplementary material Note 2. The spin Hall angle for CSHE, SAHE and MSHE can be defined by the characteristic spin Hall conductivity as $\theta_{CSH} = \sigma_{yx}^{z}/\sigma_{xx}$ , $\theta_{SAH} = \sigma_{zy}^{SAH}/\sigma_{xx}$ , $\theta_{MSH} = \sigma_{zx}^{z}/\sigma_{xx}$ and the calculated results have been shown in Fig.1. It can be seen that the conventional spin Hall angle $\theta_{CSH}$ for Fe$_{50}$Pt$_{50}$ ($\theta_{CSH}^{Fe_{50}Pt_{50}} = 0.065$) and Co$_{50}$Pt$_{50}$ ($\theta_{CSH}^{Co_{50}Pt_{50}} = 0.071$) alloys are comparable to that of Pt ($\theta_{CSH}^{Pt} = 0.08 \pm 0.02$ from experiment[3]). In addition, the magnetic spin Hall angles $\theta_{MSH}$ for Pt-based alloys are sizable ($\theta_{MSH}^{Fe_{50}Pt_{50}} = 0.044$, $\theta_{MSH}^{Co_{50}Pt_{50}} = 0.043$, $\theta_{MSH}^{Ni_{50}Pt_{50}} = -0.032$), which are at the same order of magnitude as $\theta_{CSH}$. The relative magnitude of three spin Hall angles and the main spin Hall mechanism in ferromagnetic metals depends on specific materials. For example, the leading spin Hall mechanism in $Fe_{50}Pt_{50}$ and $Co_{50}Pt_{50}$ is CSHE, followed by MSHE, and the SAHE has the smallest contribution, while for $Ni_{50}Pt_{50}$ the spin Hall angle for MSHE is largest and the order of the three spin Hall angles are $\theta_{MSH} > \theta_{CSH} > \theta_{SAH}$. The spin Hall effects in traditional 3$d$ ferromagnetic metals are also pronounced. For $Ni_{80}Fe_{20}$, it has comparable value of $\theta_{CSH}$ and $\theta_{SAH}$ but small $\theta_{MSH}$, while for $Co_{50}Fe_{50}$ the MSHE is dominating, the corresponding spin Hall angle $\theta_{MSH}$ is larger than $\theta_{CSH}$ and $\theta_{SAH}$.

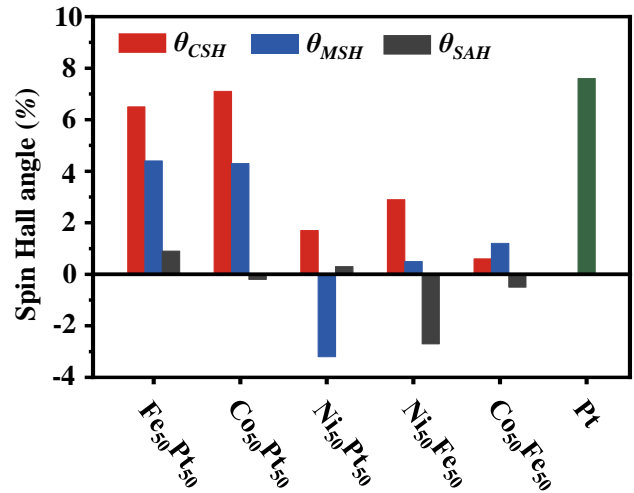

FIG. 1. The spin Hall angles corresponding to CSHE ($\theta_{CSH}$), MSHE ($\theta_{MSH}$) and SAHE ($\theta_{SAH}$) for (Fe, Co, Ni)$_{50}$Pt$_{50}$, Ni$_{80}$Fe$_{20}$ and Co$_{50}$Fe$_{50}$ alloys. The green column indicates the conventional spin Hall angle of Pt for reference.

The electronic structure properties, and therefore the relative strength of different spin Hall effect can also be optimized by varying composition of ferromagnetic alloy. As a demonstration, Fig. 2 show the spin Hall angle as a function of Fe concentration in FePt alloy (The SHCs and the longitudinal conductivity $\sigma_{xx}$ for FePt alloy can be found in supplementary material Note 3). As shown in Fig. 2 the conventional spin Hall conductivity and Hall angle decreases sharply with increasing Fe concentration.

Similar behavior also appears in SAHE. Besides, the strength of CSHE and SAHE decreases by around an order of magnitude with the increasing of Fe concentration from 25% to 75% in FePt alloy. The magnetic spin Hall angle shows a different trend with increasing Fe concentration, in which the peak of MSHE appears at $x \approx 40\%$, corresponding to $\theta_{MSH}=0.045$.

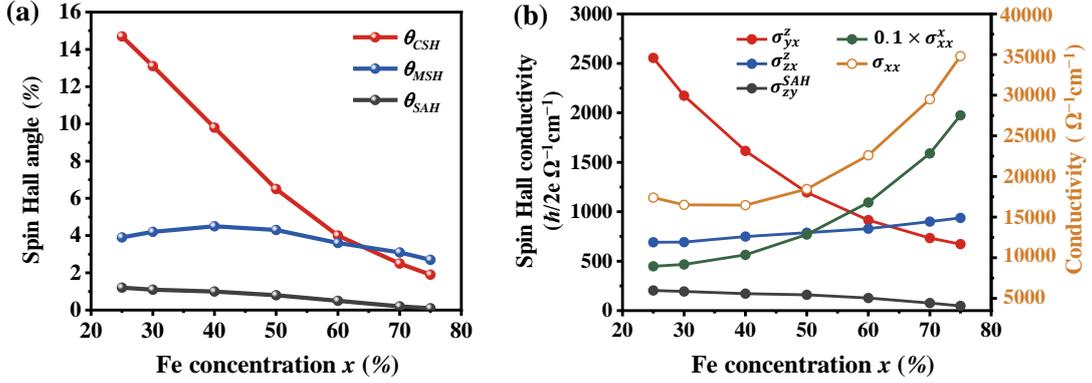

FIG. 2. The conventional, magnetic, and spin anomalous spin Hall angles(a), and spin conductivity, as well as longitudinal conductivity(b) as a function of Fe concentration in $Fe_xPt_{100-x}$ alloys.

The coexistence of significant conventional spin Hall, spin anomalous Hall, and magnetic spin Hall effects in ferromagnetic metals indicates that ferromagnets could be efficient spin-current source for generating adjustable spin-polarization and therefore, may stimulate emergent spin-charge conversion phenomena. As shown in Fig.3, let's suppose the electric current is applied along $\hat{x}$ direction in thin film of ferromagnetic metal and the direction of its magnetization $M$ can rotate in-plane. We focus on the spin current which transports along $z$-axis. By placing another ferromagnetic free layer on top of this film, this structure turns to be a typical geometry for spin Hall induced spin-orbit torque where the ferromagnetic metal shown in Fig.3 has been used as spin-current source. As shown in the figure, when the in-plane magnetization is perpendicular to the charge current, the spin current along $z$-axis will be contributed by CSHE and SAHE with its spin-polarization $\sigma$ along $y$ direction. In contrast, when the in-plane magnetization has been set to be parallel to the charge current, the spin current will be generated by CSHE as well as MSHE with its spin-polarization $\sigma$ has both $y$ and $z$ components. Importantly, the spin-polarization component $\sigma_z$ is desirable for field-free switching of perpendicularly magnetized free layer[5], which can not be generated by heavy metals Pt, Ta, *etc*. Generally, when the direction of in-plane magnetization $\hat{m}$ has $x$ and $y$ components, the spin-polarization $\sigma$ of spin current produced by all the three spin Hall effect CSHE, SAHE and MSHE, can be described as $\vec{\sigma} = [\theta_{SHC}\hat{y} + \theta_{SAHC}(\hat{m}\cdot\hat{y})\hat{m} + \theta_{MSHC}(\hat{m}\cdot\hat{x})\hat{z}]$. It's worth noting that a very recent experiment which successfully demonstrates the field-free switching of perpendicular magnetization with lower current density may suggest the presence of multiple spin-polarization of spin current generated by spin Hall effect in ferromagnetic meta[29].

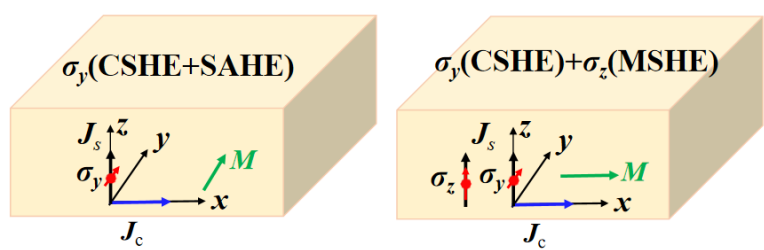

FIG.3 Illustrations of ferromagnetic metals being used as spin-current source for spin-orbit torque switching. The directions of charge current $J_c$, the magnetization $M$ and spin current $J_s$ has been marked in blue, green and black arrows, respectively. The spin-polarization $\sigma$ of the spin current has been shown in red arrow. (Left) The magnetization $M$ is perpendicular to $J_c$. (Right) The magnetization $M$ is parallel to $J_c$.

According to Onsager reciprocity, there are also inverse effects corresponding to the three spin Hall mechanism in ferromagnetic metals[21], namely iCSHE (inverse conventional spin Hall), iSAHE (inverse spin anomalous Hall), and iMSHE (inverse magnetic spin Hall). The inverse spin Hall effects in ferromagnetic metals will convert spin current to charge current, and in an open circuit, a measurable voltage. Except the spin-orbit torque application, the discovered magnetization dependant SAHE and MSHE and their inverse effect in ferromagnetic metals may also be advantageous for other spin-charge conversion related phenomena, for instance spin Seebeck and Nernst effect driven by temperature gradient[20].

At last, we briefly discuss possible experimental verification and measurement of spin Hall effect in ferromagnetic metals. In "FM1/NM/FM2" trilayer, the spin current produced by FM1 layer through spin Hall effect will produce spin-orbit torque on FM2. The effective magnetic field on FM2 applied by different spin Hall mechanism in FM1 can be determined by second harmonic Hall measurement[5], and therefore, the corresponding spin Hall angles in FM1 can be evaluate[29]. As proposed in Ref.[21], in the similar "FM1/NM/FM2" structure, the spin current in FM1 will transmit into in FM2 and lead to a Hall voltage because of the inverse spin Hall effect. The measurement on the Hall voltage may also be able to characterize the magnitude of spin Hall effect in ferromagnetic metals.

In summary, we have investigated the spin Hall effect in ferromagnetic metals and divide them into three categories including conventional, spin anomalous and magnetic spin Hall effect according to different physical mechanisms. By using first-principles calculation, we explicitly evaluate the spin Hall conductivity and spin Hall angle corresponding to three spin Hall effect for ferromagnetic alloy. For Pt-based alloy, we predict large conventional and magnetic spin Hall angle, comparable to that of nonmagnetic heavy metal. Our calculations demonstrate that, generally the three spin Hall effects coexist in ferromagnetic metals, and the relative strength depends on specific materials. As an example, we reveal that the presence of remarkable unconventional spin Hall effect in ferromagnetic metals make them efficient spin current source with out-of-plane spin-polarization for spin-orbit torque induced switching. This work provides comprehensive understanding and highlights the significance of the spin Hall effect in ferromagnetic metal, and may promote the future research progress on ferromagnet based spin-charge conversion in spintronics.

## Acknowledgment

This work was supported by the National Natural Science Foundation of China (grant No. 12174129).